\documentclass[a4paper]{article}
\usepackage{amsmath,amsthm,amssymb}
\def\vers{\mathbf{n}}
\def\de{{\rm d}}
\def\vett#1{\mathbf{#1}}
\def\vidot{\dot{\mathbf{v}}}
\def\equ{Eq.\ }

\title{On the nonradiating motion of point charges}
\author{Massimo Marino \\
{ } \\
Dipartimento di Matematica, Universit\`{a} di Milano, \\
via Saldini 50, I-20133 Milano (Italy)}

\date{July 16, 2008}

\begin{document}

\maketitle

\begin{abstract}
We investigate the possible existence of nonradiating motions of
systems of point charges, according to classical electrodynamics
with retarded potentials. We prove that two point particles of
arbitrary electric charges cannot move for an infinitely long time
within a finite region of space without radiating electromagnetic
energy. We show however with an example that nonradiating
accelerated motions of systems of point charges do in general
exist.
\end{abstract}

\section{Introduction} \label{uno}
It has already been proved since a long time (see \cite{bohm,
goed, arnettg, devaney} and the references given in \cite{goed} to
the previous works of Herglotz, Sommerfeld, Hertz and Schott) that
there exist nontrivial motions of extended electric charge
distributions which do not radiate energy according to the
classical theory of electromagnetism. Some of these motions refer
to rigid charged bodies, and were originally associated with
classical extended models of elementary particles. In more recent
times nonradiating sources have gained renewed popularity in
connection with the study of the inverse problem in wave
equations, namely, the problem of reconstructing a source when the
radiation emitted or scattered by it is known (see for instance
\cite{marengo,hoe,colton} and references therein).

Most of the existing literature in this field deals with spatially
extended sources with a monochromatic dependence on time. In the
present paper we will instead look for nonradiating systems made
of point charges, with no a priori restriction on their possible
motion. Hence our mathematical approach to the problem of
nonradiating sources will be completely different, and in some
sense complementary to that usually followed. From a fundamental
point of view, an obvious motivation for our study comes from the
fact that, according to present-day standard theories of
microscopic physics (with the exception of string theory),
elementary electric charges in nature are actually point-like.
Therefore, in these theories, continuous charge distributions can
only serve as useful approximations for the description of
macroscopic bodies.

The results of the present investigation may be relevant in
connection with the search for classical models of atomic systems.
A classical atom is in fact usually described as an isolated
system of moving point charges (nucleus and electrons). Since the
absence of radiation is a necessary condition for the stability of
the atomic ground state, the formulation of a classical atomic
model can only be possible provided that nonradiating motions of
point charges indeed exist, and that they are compatible with
suitable dynamical laws taking into account radiation reaction
(one may adopt for instance the classical third order
Lorentz--Dirac equation \cite{lorentz, dirac}, or its approximated
version of second order with respect to time \cite{ll, spohn}).
There is a widespread belief that all solutions of these dynamical
equations do actually radiate, and that classical physics
therefore cannot account for the stability of atomic systems.
Nevertheless, attempts to describe atomic physics by making only
use of the laws of classical electrodynamics have still recently
been undertaken \cite{deluca}. Moreover, nonradiating motions have
been found in the dipole approximation for infinite regular arrays
(with arbitrary lattice parameter) of point-like charged
oscillators obeying to the Lorentz--Dirac equation with retarded
mutual electromagnetic interactions \cite{cg, mcg}. It would
therefore be interesting, in our opinion, to establish in which
cases the impossibility of nonradiating motions of a finite number
of point charges can indeed be proved with rigorous mathematical
arguments.

In this work we shall not be directly concerned with the dynamical
equations which determine the motion of the particles when the
field acting on them is known. Our aim will be simply to study the
restrictions which are imposed on any arbitrary motion of point
charges by the condition of vanishing radiation. This radiation
will be calculated by making use of the usual retarded
electromagnetic potentials which arise from Maxwell equations with
point-like sources. Particles moving with constant velocities of
course do not radiate electromagnetic energy, and their distance
from any fixed point in space increases to infinity whenever their
velocities are not zero. For the case of one or two particles with
arbitrary charges, we will prove that there do not exist
nonradiating motions which are bounded in space for all times.
However, we will find that a pair of point particles of equal
charges do not radiate while they move on a straight line in
opposite directions, with their spatial coordinates varying as the
square root of time. This is an indication of the fact that
accelerated nonradiating motions of systems of point charges do in
general exist. The search for other nontrivial examples involving
a number of particles greater than two could be an interesting
matter for further investigation.

\section{The condition of zero radiation}\label{due}

Let us formulate our problem in more precise mathematical terms.
We shall consider smooth motions of $N$ point particles, with
nonvanishing electric charges $q_1, \dots, q_N$ of arbitrary
magnitudes and signs. We denote by the three-dimensional vector
$\vett{z}_i(t)$ the position of particle $i$ at time $t$, with
respect to a fixed cartesian coordinate system. We suppose that
the motion of these charges is confined within a finite region of
space. This means that there exists a fixed length $L>0$ such that
\begin{equation}\label{limi}
|\vett{z}_i(t)|<L \qquad \text{for all $t$ and all } i=1, \dots,
N\,.
\end{equation}
The condition of no radiation is expressed as the requirement that
the flux of the Poynting vector generated by the charges,
calculated through any large spherical surface of radius $R\gg L$,
vanishes at all times. Apart from the conditions just mentioned,
the motions considered will be a priori of the most general
possible type.

The retarded electric field generated at space-time point
$(\vett{x},t)$ by point particle $i$ is \cite{jackson}
\begin{equation} \label{elec}
\mathbf{E}_i(\vett{x},t)= \frac {q_i}{R_i(1- \vers_i\cdot
\vett{v}_i)^2} \left[-\vidot_i+ \left(\vers_i\cdot\vidot_i+
\frac{1-v_i^2}{R_i}\right) \frac {\vers_i- \vett{v}_i} {1-
\vers_i\cdot \vett{v}_i} \right]\,.
\end{equation}
Here $R_i=|\vett{x}-\vett{z}_i(t_i)|$ and $\vers_i= [\vett{x}-
\vett{z}_i(t_i)]/R_i$ (so that $|\vers_i|=1$), where $t_i$ is the
retarded time of particle $i$, which is defined as an implicit
function of $(\vett{x},t)$ by the equation
\begin{equation}\label{retard}
t-t_i=|\vett{x}-\vett{z}_i(t_i)|\,.
\end{equation}
In (\ref{elec}) we put also $\vett{v}_i= \vett{v}_i(t_i)=
\de\vett{z}_i(t_i)/\de t_i$ and $\vidot_i= \vidot_i(t_i)=
\de\vett{v}_i(t_i)/\de t_i$. The magnetic field can then be
expressed as
\begin{equation}\label{magn}
\mathbf{B}_i(\vett{x},t)= \vers_i \times \mathbf{E}_i
(\vett{x},t)\,.
\end{equation}
Note that in our units the speed of light is 1.

Let us evaluate the retarded field (\ref{elec}) at a point
$\vett{x}$ such that $|\vett{x}|=R$. In the limit $R\to \infty$ we
have $\vers_i= \vers +O(R^{-1})$, where the unit vector $\vers=
\vett{x}/R$ defines the particular direction considered. We also
have $R_i= R-\vers\cdot \vett{z}_i(t_i) + O(R^{-1})$. Therefore,
if we call $t_R$ the time at which the field is evaluated at
$\vett{x}$, using \equ (\ref{retard}) and neglecting infinitesimal
terms we can write
\begin{equation}
t_i=t+\vers\cdot\mathbf{z}_i(t_i)\,, \label{tconi}
\end{equation}
where $t=t_R -R$. It follows that for $R\to \infty$ \equ
(\ref{elec}) can be simplified as
\begin{equation} \label{elec1}
\mathbf{E}_i(\vett{x},t_R)= \frac {q_i}{R(1- \vers\cdot
\vett{v}_i)^2} \left[-\vidot_i+ \frac {\vers \cdot\vidot_i} {1-
\vers\cdot \vett{v}_i} (\vers- \vett{v}_i)\right]+ O(R^{-2})\,,
\end{equation}
where $\vett{v}_i$ and $\vidot_i$ are evaluated at the time $t_i$
which is implicitly defined by \equ (\ref{tconi}). It is
convenient for our purposes that in this equation $t$ be
considered independent of $R$. This means that the retarded time
$t_i$ is also independent of $R$, whereas $t_R=t+ R$ must increase
with $R$ for fixed $t$.

For the total fields generated by the system of particles we have
$\vett{E}= R^{-1}\bar{\vett{E}}+ O(R^{-2})$, $\vett{B}=
R^{-1}\vers\times \bar{\vett{E}}+ O(R^{-2})$, where
\begin{equation}
\bar{\vett{E}}(\vers, t)= \sum_{i=1}^{N} \frac {q_i}{(1-
\vers\cdot \vett{v}_i)^2} \left[-\vidot_i+ \frac {\vers
\cdot\vidot_i} {1- \vers\cdot \vett{v}_i} (\vers-
\vett{v}_i)\right]
\end{equation}
is a quantity independent of $R$. The power radiated by the system
at the time $t$ can be defined as the flux $\Phi$ of the Poynting
vector $\vett{S}= (1/4\pi) \vett{E} \times \vett{B}$ through a
sphere $\Sigma$ of radius $R$ at the time $t_R$, in the limit of
large $R$. We have
\begin{equation}
\Phi=\lim_{R\to \infty}R^2 \int \de\Omega_\vers\, \vett{S}\cdot
\vers= \int \frac{\de\Omega_\vers}{4\pi} [\bar{\vett{E}}\times
(\vers\times \bar{\vett{E}})]\cdot \vers = \int
\frac{\de\Omega_\vers}{4\pi} (\vers\times \bar{\vett{E}})^2 \,,
\end{equation}
where $\Omega_\vers$ denotes the solid angle associated with the
direction $\vers$, and integration is carried out over the total
solid angle. Therefore the condition $\Phi=0$ is equivalent to the
requirement that the vector $\vers\times \bar{\vett{E}}$ vanishes
for all directions $\vers$ and times $t$:
\begin{equation}\label{nr}
0=\vers\times \bar{\vett{E}}= -\vers\times \sum_{i=1}^{N} \frac
{q_i}{(1- \vers\cdot \vett{v}_i)^2} \left(\vidot_i+ \frac {\vers
\cdot\vidot_i} {1- \vers\cdot \vett{v}_i} \vett{v}_i\right)\,.
\end{equation}

The right-hand side of the above equation can be rewritten in a
particularly compact form. In fact, using (\ref{tconi}) we find
that the partial derivative of $t_i$ with respect to $t$ at fixed
$\vers$ is
\begin{equation}\label{pd}
\left.\frac{\partial t_i}{\partial t}\right|_{\vers} =\frac
1{1-\mathbf{v}_i(t_i)\cdot\vers}\,.
\end{equation}
We then have
\begin{equation}
\left.\frac{\partial \vett{z}_i}{\partial t}\right|_{\vers} =\frac
{\vett{v}_i}{1-\mathbf{v}_i\cdot\vers}
\end{equation}
and
\begin{equation}
\left.\frac{\partial^2 \vett{z}_i}{\partial t^2}\right|_{\vers}
=\frac {1}{(1- \vers\cdot \vett{v}_i)^2} \left(\vidot_i+ \frac
{\vers \cdot\vidot_i} {1- \vers\cdot \vett{v}_i}
\vett{v}_i\right)\,,
\end{equation}
where all quantities with label $i$ are evaluated at the time
$t_i$. Hence (\ref{nr}) can be rewritten as
\begin{equation}\label{dt}
\vers\times \sum_{i=1}^N q_i\left.\frac{\partial^2
\vett{z}_i}{\partial t^2}\right|_{\vers}=0\,.
\end{equation}
Equation (\ref{tconi}) implies that
\begin{equation}\label{som}
\sum_{i=1}^N q_i\vett{z}_i(t_i)= \sum_{i=1}^N q_i(t_i-t)\vers+
\vett{C}(\vers,t)\,,
\end{equation}
where $\vett{C}(\vers,t)$ is an arbitrary function such that
$\vett{C}(\vers,t) \cdot \vers=0$. Substituting the above
expression into (\ref{dt}) we then obtain
\begin{equation}
\frac{\partial^2}{\partial t^2} \vett{C}(\vers,t)=0\,,
\end{equation}
so that
\begin{equation}\label{c}
\vett{C}(\vers,t)= \vett{C}_0 (\vers)+ t\vett{C}_1 (\vers)\,,
\end{equation}
where $\vett{C}_0 (\vers)$ and $\vett{C}_1 (\vers)$ are functions
defined on the unit sphere $|\vers|=1$, such that $\vett{C}_0
(\vers)\cdot\vers= \vett{C}_1 (\vers)\cdot\vers= 0$. According to
(\ref{limi}), we must have
\begin{equation}\label{bnd}
\left|\sum_{i=1}^N q_i\vett{z}_i(t_i)\right| <L \sum_{i=1}^N |q_i|
\end{equation}
for all $t$. On the other hand, from (\ref{som}) and (\ref{c}) we
obtain
\[
\left|\sum_{i=1}^N q_i\vett{z}_i(t_i)\right| \geq |\vett{C}_0
(\vers)+ t\vett{C}_1 (\vers)|\,.
\]
Therefore condition (\ref{bnd}) can be satisfied for $t\to\infty$
only provided that $\vett{C}_1 (\vers)=0$ for all $\vers$. We then
conclude that
\begin{equation}\label{som1}
\sum_{i=1}^N q_i[\vett{z}_i(t_i)-(t_i-t)\vers]= \vett{C}_0
(\vers)\,,
\end{equation}
with $\vett{C}_0 (\vers)\cdot \vers =0$. We can express this
result by saying that a bounded motion of a system of $N$ charges
does not radiate electromagnetic energy if and only if the
quantity on the left-hand side of (\ref{som1}), where $t_i$ is
determined by \equ (\ref{tconi}) for all $i=1, \dots, N$, is
independent of $t$ for all unit vectors $\vers$.

By differentiating (\ref{som1}) with respect to $t$ and using
(\ref{pd}) we obtain
\begin{equation}\label{mc}
\sum_{i=1}^N q_i \frac{\vett{v}_i(t_i)- [\vers\cdot
\vett{v}_i(t_i)]\vers} {1-\vers\cdot\vett{v}_i(t_i)} =0 \,.
\end{equation}
From the above formula it is easy to recover the well-known result
that a single charged particle moving in a bounded region of space
necessarily radiates. For $N=1$ in fact (\ref{mc}) is equivalent
to $\vett{v}(t)= [\vers\cdot \vett{v}(t)]\vers$, which means that
$\vett{v}(t)$ must be parallel to $\vers$. Since $\vers$ can be
varied independently of $t$, this condition implies
$\vett{v}(t)=0$. Hence the particle must necessarily be static for
all times in order to satisfy the condition of zero radiation.

\section{The system of two charges}\label{tre}

Let us consider the case $N=2$. Condition (\ref{mc}) of zero
radiation can be written as
\begin{equation}\label{bigv}
\vett{V}= (\vers\cdot \vett{V})\vers\,,
\end{equation}
where
\begin{equation}\label{bigv2}
\vett{V}= q_1[1-\vers\cdot\vett{v}_2(t_2)]\vett{v}_1 (t_1)+ q_2[1-
\vers\cdot\vett{v}_1(t_1)] \vett{v}_2(t_2) \,.
\end{equation}
In the above equation the times $t_1$ and $t_2$ are determined by
\equ (\ref{tconi}) for $i=1,2$. Hence we have
\begin{equation}\label{effe}
t_2-t_1= [\mathbf{z}_2(t_2)-\mathbf{z}_1(t_1)] \cdot\vers \,,
\end{equation}
which implies
\begin{equation}\label{effe2}
|t_2-t_1| <|\mathbf{z}_2(t_2)-\mathbf{z}_1(t_1)| \,.
\end{equation}
In relativistic language, the above relation means that the two
spacetime points $(\vett{z}_1,t_1)$ and $(\vett{z}_2,t_2)$, taken
on the world-lines of particle 1 and 2 respectively, must have a
space-like separation. Clearly, for any $t_1$ such that
$\mathbf{z}_2(t_1) \neq \mathbf{z}_1(t_1)$, there exists a finite
interval of values of $t_2$, including the point $t_2=t_1$, for
which (\ref{effe2}) is satisfied. Note that, if $\vett{z}_1(t)=
\vett{z}_2(t)$ for all $t$, then the two particles actually form a
single compound particle with charge $q_1+q_2$, so that the
situation is identical to the case of a single charge, which has
already been considered at the end of the preceding section.
Therefore, excluding this trivial case, in the following we shall
always assume that $\vett{z}_1(t)\neq \vett{z}_2(t)$ for almost
all $t$.

Taking into account the arbitrariness of $\vers$ and $t$, we see
from (\ref{effe}) that, if one takes any two times $t_1$ and $t_2$
satisfying (\ref{effe2}), then (\ref{bigv}) must hold for all unit
vectors $\vers$ forming with $\mathbf{z}_2(t_2)
-\mathbf{z}_1(t_1)$ the angle
\begin{equation}\label{effe1}
\theta= \arccos\frac{t_2-t_1}
{|\mathbf{z}_2(t_2)-\mathbf{z}_1(t_1)|} \,.
\end{equation}
The set of all such unit vectors forms a circle $C_\theta$ of
radius $\sin\theta$ on the unit sphere. In order to satisfy \equ
(\ref{bigv}), $\vett{V}$ has to be parallel to $\vers$ for all
$\vers\in C_\theta$. However, for any $\theta\neq \pi/2$, the
circle $C_\theta$ is not contained in any plane containing the
origin of the cartesian system. This implies, in particular, that
almost all $\vers\in C_\theta$ do not lie in the plane containing
$\vett{v}_1(t_1)$ and $\vett{v}_2(t_2)$. Since $\vett{V}$ lies
instead in this plane for any $\vers\in C_\theta$, we see that
\equ (\ref{bigv}) can be satisfied only if $\vett{V}=0$. From
(\ref{bigv2}) it then follows that $\vett{v}_2(t_2)$ must be
parallel to $\vett{v}_1(t_1)$. Substituting $\vett{v}_2(t_2)=
\lambda \vett{v}_1(t_1)$ into the equation $\vett{V}=0$, and
solving with respect to $\lambda$, we obtain
\begin{equation}
\vett{v}_2(t_2) = -\frac{q_1 \vett{v}_1(t_1)}{q_2- (q_1+q_2)
\vers\cdot\vett{v}_1(t_1)}\,. \label{s2v2}
\end{equation}

Let us initially suppose that $q_1+q_2=0$, which means that we are
dealing with a neutral two-particle system (such as a hydrogen
atom). Then (\ref{s2v2}) becomes
\begin{equation}\label{vv}
\mathbf{v}_1(t_1) =\mathbf{v}_2(t_2) 
\end{equation}
for all $t_1$ and $t_2$ satisfying (\ref{effe2}). This means that,
if $t$ is such that $\mathbf{z}_2(t)\neq \mathbf{z}_1(t)$, then
(\ref{vv}) is satisfied for $t_1=t$ and for all $t_2$ belonging to
a finite interval containing $t$. This implies in particular that
$\de\mathbf{v}_2(t_2)/\de t_2 =0$ for $t_2=t$. In the same way, by
interchanging the role of particles 1 and 2, we also obtain that
$\de\mathbf{v}_1(t_1)/\de t_1=0$ for $t_1=t$. We have thus proved
that $\mathbf{v}_1(t)= \mathbf{v}_2(t)$ and
$\dot{\mathbf{v}}_1(t)= \dot{\mathbf{v}}_2(t)= 0$ for all $t$ such
that $\mathbf{z}_2(t) \neq \mathbf{z}_1(t)$. From this fact it
easily follows that $\mathbf{v}_1(t)=\mathbf{v}_2(t)=$ constant
for all $t$. Since the trajectories of the two particles were
supposed to be bounded in space, we then conclude that
\begin{equation}
\mathbf{v}_1(t)=\mathbf{v}_2(t)=0 \quad\text {for all }t\,.
\end{equation}
Therefore, for two charged particles such that $q_1+ q_2=0$, the
radiated power vanishes at all times only if the particles are
static.

Let us now consider the case $q_1+ q_2\neq 0$. If we take any two
times $t_1$ and $t_2$ satisfying (\ref{effe2}), we see from
(\ref{s2v2}) and (\ref{effe1}) that $\vers\cdot \vett{v}_1(t_1)$
must be a constant while $\vers$ varies in the circle $C_\theta$.
This implies that $\vett{v}_1(t_1)$ must be directed as
$\mathbf{z}_2(t_2) -\mathbf{z}_1(t_1)$. By keeping $t_2$ fixed and
varying $\theta$, one can actually prove that this fact is true
for all $t_1$ belonging to a finite interval of time. A
symmetrical result can also be proved for $\vett{v}_2(t_2)$. We
thus conclude that the whole motion of both particles must take
place along a straight line. We have therefore reduced the problem
to the study of a one-dimensional motion, and we will henceforth
denote as $z_1$ and $z_2$ the (scalar) coordinates of the two
particles. We can rewrite (\ref{effe1}) as
\begin{equation}\label{effe3}
[z_2(t_2)-z_1(t_1)] \cos\theta = t_2-t_1 \,,
\end{equation}
and we have
\[
\vers\cdot\vett{v}_1(t_1)= v_1(t_1) \cos\theta= v_1(t_1)
\frac{t_2-t_1} {z_2(t_2)- z_1(t_1)}\,,
\]
with $v_1(t_1)=\de z_1/\de t_1$, $v_2(t_2)=\de z_2/\de t_2$.
Therefore from (\ref{s2v2}) we obtain
\begin{equation}\label{sv}
[z_2(t_2)-z_1(t_1)][q_1 v_1(t_1) +q_2
v_2(t_2)]-(q_1+q_2)(t_2-t_1)v_1(t_1) v_2(t_2)=0\,.
\end{equation}
This equation must hold for all $t_1$ and $t_2$ satisfying \equ
(\ref{effe2}). By setting $t_1=t_2=t$, where $t$ is such that
$z_1(t)\neq z_2(t)$, we obtain
\[
0=q_1 v_1(t)+ q_2 v_2(t)= \frac \de{\de t}[q_1 z_1(t)+ q_2 z_2(t)]
\,,
\]
which means that $q_1 z_1(t)+ q_2 z_2(t)$ is a constant. Since
$q_1+ q_2 \neq 0$, by suitably choosing the origin of the $z$ axis
we can always set this constant to 0 and obtain
\begin{equation}\label{z2}
z_2(t)=-\frac{q_1}{q_2}z_1(t)\,, \quad v_2(t)= -\frac{q_1}{q_2}
v_1(t) \quad \text{for all }t\,.
\end{equation}
Hence, writing $z$ and $v$ in place of $z_1$ and $v_1$
respectively, \equ (\ref{sv}) becomes
\begin{equation} \label{zv}
[q_1 z(t_2)+ q_2 z(t_1)] [v(t_2) -v(t_1)] +(q_1+ q_2)(t_2-
t_1)v(t_1)v(t_2) =0 \,.
\end{equation}
According to (\ref{effe2}) and (\ref{z2}), this equality must be
true for all $t_1$ and $t_2$ such that
\begin{equation}\label{effe4}
|q_2(t_2-t_1)| <|q_1 z(t_2)+ q_2 z(t_1)| \,.
\end{equation}
Interchanging $t_1$ and $t_2$ in \equ (\ref{zv}), we also get
\begin{equation} \label{zv2}
[q_1 z(t_1)+ q_2 z(t_2)] [v(t_2) -v(t_1)] +(q_1+ q_2)(t_2
-t_1)v(t_1)v(t_2) =0 
\end{equation}
for
\begin{equation}\label{effe5}
|q_2(t_2-t_1)| <|q_1 z(t_1)+ q_2 z(t_2)| \,.
\end{equation}
For all $t_1$ and $t_2$ such that $|t_2-t_1|$ is sufficiently
small, both conditions (\ref{effe4}) and (\ref{effe5}) are
simultaneously satisfied. Therefore, subtracting (\ref{zv2}) from
(\ref{zv}) we obtain
\[
(q_2-q_1)[z(t_2)-z(t_1)][v(t_2)-v(t_1)]=0 \,.
\]
If $q_1\neq q_2$, the above equation implies that $v(t_1)=v(t_2)$
for all $t_1$ and $t_2$ such that $z(t_1)\neq z(t_2)$. But for
such $t_1$ and $t_2$ then (\ref{zv}) implies that
$v(t_1)=v(t_2)=0$. On the other hand, for any regular function
$z(t)$, if $z(t_1)\neq z(t_2)$ there must be a time $\bar t$
between $t_1$ and $t_2$, such that $z(\bar t)\neq z(t_1)$ and
$v(\bar t) =\dot z(\bar t) \neq 0$. We see therefore that the
hypothesis $z(t_1)\neq z(t_2)$ leads to a contradiction. We must
thus have $z(t_1)= z(t_2)$ for all $t_1$ and $t_2$, which means
that the particles are motionless.

Let us finally suppose that $q_1=q_2$. Then (\ref{zv}) becomes
\begin{equation} \label{zv3}
[z(t_2)+ z(t_1)] [v(t_2) -v(t_1)] +2(t_2- t_1)v(t_1)v(t_2) =0 \,,
\end{equation}
with $z(t)=z_1(t)=-z_2(t)$, $v(t)=v_1(t)=-v_2(t)$. Let us divide
(\ref{zv3}) by $t_2- t_1$ and take the limit for $t_2\to t_1 =t$.
We obtain
\[
\ddot z (t)z(t)+ \dot z^2(t)=0\,,
\]
or
\[
\frac {\de^2}{\de t^2}z^2(t)=0\,.
\]
By integrating this equation we get
\begin{equation}\label{nt}
z(t)= \sqrt{a+bt}\,,
\end{equation}
where $a$ and $b$ are two integration constants. It is interesting
to observe that this function $z(t)$ is indeed a solution of \equ
(\ref{zv3}) for all $a$ and $b$, so that it really describes a
nonradiating motion of two equal charges. If $b \neq 0$, by a
suitable shift of the time axis it is always possible to set $a=0$
in (\ref{nt}). Then the equations
\begin{equation}\label{nt2}
z_1(t)=\sqrt{bt}\,, \qquad z_2(t)=-\sqrt{bt}
\end{equation}
represent a nonradiating motion which is defined for all $t$ such
that $bt \geq 0$. If we also consider the conditions $|v_1(t)|<1$,
$|v_2(t)|<1$, which are imposed by special relativity, we must
require that $(1/2)\sqrt{|b/t|}<1$, or $|t|> |b|/4$. The motion is
thus physically meaningful in the time interval $b/4<t< +\infty$
if $b>0$, or $-\infty < t<b/4$ if $b<0$. However, in order for the
motion to be bounded we must necessarily have $b=0$ in formula
(\ref{nt}), so that we again conclude that the particles must be
motionless. This result definitely excludes the possibility of a
nonradiating bounded motion of any pair of arbitrary point
charges.

\section{Discussion}

We have obtained a general condition for the absence of
electromagnetic radiation from a system of moving point charges.
This condition, which is expressed by formula (\ref{mc}), seems at
first sight very restrictive, since it must be satisfied for any
arbitrary direction $\vers$. Using it we have deduced that two
point particles of arbitrary electric charges cannot move for an
infinitely long time within a finite region of space without
radiating electromagnetic energy. However, an analogous result for
more than two charged particles is at present not available. We
have shown on the other hand that, if one only considers finite
intervals of time (or, conversely, if one takes also unbounded
trajectories into consideration), then nontrivial nonradiating
motions of systems of point charges actually exist: \equ
(\ref{nt2}) provides an example of such a motion for $q_1=q_2$.

In classical electrodynamics it is possible to remove in a
relativistically covariant way the divergences which are
associated with the presence of point charges, and to obtain
finite expressions for the energy and momentum of the complete
system of particles and field \cite{marino1, lech}. The
conservation of these ``renormalized'' quantities imposes on the
particles the Lorentz--Dirac equation of motion. However, the
renormalized electromagnetic energy in the presence of point
charges is no longer a positive definite functional of the field
configuration \cite{marino1}. Therefore a spatially confined
system of point charges might in principle keep radiating for an
infinitely long time, while the electromagnetic energy contained
in a finite volume including the particles diverges toward $-
\infty$. For an isolated system, the electromagnetic field can be
entirely expressed as a function of the dynamical variables of the
particles at retarded times. Hence the divergence of the
electromagnetic energy with increasing time must in any case be
associated with an irreversible behavior of the system. It follows
that a physically acceptable description of a stable system, such
as an atom in its ground state, requires the existence of
solutions which do not radiate, or which radiate at most a finite
amount of energy during their whole history, starting from a given
initial time.

Let us consider a hypothetical confined solution of the
Lorentz--Dirac equation for two interacting particles with charges
of equal modulus and opposite sign. Suppose also that this
solution is such that the particles do not fall into each other
either at finite or infinite times. Then the results of the
preceding section suggest that, in order for the system to radiate
at most a finite amount of energy, the accelerations of the two
particles must tend asymptotically to zero. However such a motion
is obviously not a solution of the Lorentz--Dirac equation, since
the Coulomb attractive force does not asymptotically vanish. We
conclude that the description of the hydrogen atom as an isolated
system governed by the laws of classical electromagnetism is
incapable of accounting for the existence of bound noncollapsing
states.

It is well known that, if radiation reaction is treated as a small
perturbation of the mechanical trajectories for a charged particle
in a Coulomb field, then a particle in a bound state should spiral
toward the center of force and ultimately fall into it. At
variance with the nonrelativistic case, the total energy radiated
during such a process appears to be finite according to
relativistic mechanics \cite{boyer}. The situation becomes however
completely different if one treats the Coulomb problem by making
use of the Lorentz--Dirac equation in an exact way. It has in fact
been proved, either in the one-dimensional relativistic case
\cite{eliezer} or in the three-dimensional nonrelativistic case
\cite{carati}, that there exists no solution of the Lorentz--Dirac
equation for which the particle falls into the fixed center of
force either at finite or infinite times. An analogous result,
with not rigorous but quite convincing arguments, has also been
obtained in the relativistic three-dimensional case
\cite{eliezer}. It has also been shown in \cite{eliezer} that,
according to the relativistic Lorentz--Dirac equation, no
collision can occur between two interacting particles of equal
masses and opposite charges moving on a straight line. Let us now
make the plausible hypothesis that these results can be extended
to the case of two particles of different masses moving in
three-dimensional space. In other words, let us suppose that for
two particles there exist no collapsing solution at all. Since we
have shown that nonradiating confined solutions of the
Lorentz--Dirac equation do not exist, a confined solution should
necessarily radiate an infinite amount of energy for infinite
times. Hence the energy in a finite volume containing the system
should diverge to $-\infty$. Although we are unable at the moment
to mathematically prove the impossibility of a noncollapsing
solution of this type, its existence would be quite surprising
from a physical point of view. Taking all these facts into
consideration, we are led to make the conjecture that, for the
electromagnetic two-body problem with particles obeying to the
Lorentz--Dirac equation, the only possible solutions are given by
unbounded orbits. It is interesting in this respect to recall
that, according to a recently obtained result \cite{marino2}, only
unbounded orbits can exist for a particle in a Coulomb field in
the three-dimensional nonrelativistic case.

\end{document}